\documentclass[aps,prd,reprint,nofootinbib,superscriptaddress,longbibliography]{revtex4-2}

\usepackage{amsmath,amssymb,bm}
\usepackage{graphicx}
\usepackage{booktabs}
\usepackage{xcolor}
\usepackage[colorlinks=true,citecolor=blue,linkcolor=blue,urlcolor=blue]{hyperref}

\graphicspath{{figures/}}

\newcommand{\dd}{\mathrm{d}}
\newcommand{\ii}{\mathrm{i}}

\newcommand{\kh}{\hat{\mathbf k}}

\newcommand{\Tr}{\mathrm{Tr}}
\newcommand{\Kern}{\mathcal K}
\newcommand{\Trans}{\mathcal T}
\newcommand{\Om}{\Omega_{\rm GW}}
\begin{document}

\title{Cross-frequency SGWB anisotropy from compact topology: CMB B-mode covariance as a transfer probe}

\author{Li-e Qiang}
\email{qianglie@nssc.ac.cn}
\affiliation{National Space Science Center, Chinese Academy of Sciences, Beijing 100190, China}

\author{Peng Xu}
\email{xupeng@imech.ac.cn}
\affiliation{Center for Gravitational Wave Experiment, National Microgravity Laboratory, Institute of Mechanics, Chinese Academy of Sciences, Beijing 100190, China}
\affiliation{Key Laboratory of Gravitational Wave Precision Measurement of Zhejiang Province, Hangzhou Institute for Advanced Study, UCAS, Hangzhou 310024, China}
\affiliation{Taiji Laboratory for Gravitational Wave Universe (Beijing/Hangzhou), University of Chinese Academy of Sciences (UCAS), Beijing 100049, China}

\begin{abstract}
Compact spatial topology restricts the eigenmodes of primordial tensor perturbations, and the resulting discreteness can render the primordial stochastic gravitational-wave background (SGWB) anisotropic.  Here we treat the CMB tensor $B$-mode covariance as a transfer-filtered measurement of that ultra-low-frequency anisotropy.  Writing the normalized angular tensor-power measure as $F(k,\hat k)=1+Q(k,\hat k)$ and its nonmonopole moments as $q_{LM}(k)$, we obtain an explicit kernel that maps $q_{LM}(k)$ onto the off-diagonal covariance $\delta C^{BB}_{\ell m,\ell' m'}$.  The kernel factorizes into tensor transfer functions and a spin-weighted Gaunt coefficient and obeys the parity rule $L+\ell+\ell'$ even for $BB$ and odd for $TB/EB$.  It is an exact source--response representation of the full compact covariance rather than an additional observable.  For a cubic three-torus the geometry pins down a common cubic angular subspace and orientation across frequency bands, although the amplitudes of the allowed multipoles still depend on the radial shell and source spectrum.  The same topology-restricted template can therefore be read out either through the CMB $B$-mode kernel or through the anisotropy response of PTA/LISA/Taiji/TianQin searches.  Using CAMB transfer functions and an invariant anisotropic-template statistic, we contrast this tensor channel with the scalar $T/E$ covariance.  Independent direct angular-shell sums and $q_{LM}$--Gaunt contractions agree at $L_q^{\max}=2\ell_{\max}$ to relative Frobenius residuals of $1.4\times10^{-14}$--$3.0\times10^{-14}$.  The scalar sector holds most of the practical CMB topology information; a fixed-template scan places the combined full-sky $S/N=1$ transition between $L/\chi_*=2.34$ and $2.36$, while the $B$-mode channel remains subthreshold but isolates the primordial SGWB contribution.  We use these results to set out a cross-frequency template for future topology searches.
\end{abstract}

\maketitle

\section{Introduction}

The global topology of space is not determined by local curvature.  A Friedmann-Lemaitre-Robertson-Walker universe may have compact spatial sections, such as a cubic three-torus $T^3$ or a three-sphere $S^3$, without changing the local Einstein equations \cite{LachiezeRey1995,Levin2002}.  Matched circles arise in the universal covering space when the observer's last-scattering sphere intersects the corresponding sphere centered on a topological image of the observer \cite{Cornish1998,PlanckTopology2016,Petersen2023}.  Such matched circles are absent once the relevant observer--image separations exceed the last-scattering diameter.  Information can nevertheless survive in the covariance: compact boundary conditions break continuous statistical isotropy and correlate different $(\ell,m)$ multipoles, motivating covariance-based topology searches \cite{Akrami2024,Eskilt2024,Samandar2025}.

General maps from direction-dependent primordial tensor power to off-diagonal CMB $T/E/B$ covariance have been developed previously \cite{Shiraishi2014,Hiramatsu2018}.  COMPACT Part IIIa showed how nontrivial topology can generate parity-odd CMB correlations without parity-violating microphysics \cite{Samandar2024PartIIIa}, while Part IIIb provided spin-2 eigenmodes and complete tensor-induced CMB covariance matrices for orientable Euclidean manifolds \cite{Samandar2025}.  Here we isolate the topology-dependent SGWB source multipoles and their response.  Compact tensor eigenmodes make the primordial SGWB power anisotropic, and the CMB tensor $B$-mode covariance probes that anisotropy through the recombination and reionization transfer functions.  The full covariance remains the endpoint statistic; our contribution is an exact source--response factorization that exposes topology-constrained $q_{LM}(k)$ coordinates and their reuse with CMB and direct-SGWB response maps.

The relevant mapping is
\begin{equation*}
 \begin{gathered}
 \hbox{compact tensor eigenmodes}\;\longrightarrow\; q_{LM}(k)\\
 \longrightarrow\; \delta C^{BB}_{\ell m,\ell' m'} .
 \end{gathered}
\end{equation*}
The CMB is not a PTA- or interferometer-band SGWB detector.  It samples the same primordial tensor two-point function at the ultra-low effective wave numbers $k\sim\ell/\chi_*$ set by recombination and reionization transfer functions \cite{ZaldarriagaSeljak1997,Kamionkowski1997,PritchardKamionkowski2005}.  Direct SGWB searches with PTAs and space interferometers instead reconstruct an angular SGWB map through detector response functions \cite{RomanoCornish2017,Mingarelli2013,TaylorGair2013,Agazie2023Anisotropy,Bartolo2022LISAAnisotropy,ZhaoWang2024TaijiLISA}.  Compact topology links the two: it supplies a shared geometric source space that both kinds of measurement sample.

The remainder of the Letter derives the SGWB-anisotropy--to--$B$-mode covariance kernel, evaluates its source multipoles for a cubic $T^3$, develops the cross-frequency interpretation, and reports direct-sum validation and feasibility diagnostics.  Throughout, scalar $T/E$ covariance remains the dominant practical CMB topology channel, and the $B$-mode kernel isolates the primordial SGWB part.

\section{Primordial SGWB anisotropy}

Consider an unpolarized, parity-even primordial tensor background in the covering space.  We use $\lambda=\pm1$
as a helicity-sign label; the corresponding physical tensor helicity is $h=2\lambda=\pm2$.  Thus ${}_{-2\lambda}Y_{\ell m}={}_{-h}Y_{\ell m}$ and the $B$-mode phase convention below, $\epsilon_B^\lambda=\ii\lambda$, is equivalently $\ii h/2$.  Write
\begin{equation}
 h_{ij}(\eta,\mathbf x)=\sum_{\lambda=\pm1}\int\frac{\dd^3 k}{(2\pi)^3}
 h_\lambda(\mathbf k)e^{\lambda}_{ij}(\kh)T_h(k,\eta)e^{\ii\mathbf k\cdot\mathbf x} .
 \label{eq:tensorfield}
\end{equation}
Allow statistical homogeneity but not directional isotropy:
\begin{align}
 \left\langle h_\lambda(\mathbf k)h_{\lambda'}^*(\mathbf k')\right\rangle
 &= (2\pi)^3\delta^{(3)}(\mathbf k-\mathbf k')\delta_{\lambda\lambda'}\nonumber\\
 &\quad\times\frac{2\pi^2}{k^3}\frac{P_h(k)}{2}
 \left[1+Q(k,\kh)\right],
 \label{eq:tensor-stat}
\end{align}
where $P_h$ is the polarization-summed dimensionless tensor power.  The explicit factor $1/2$ assigns equal power to the two helicities.

It is useful to distinguish the normalized angular spectral measure
\begin{equation}
 F(k,\kh)\equiv1+Q(k,\kh),\qquad
 \frac{1}{4\pi}\int\dd\Omega_{\kh}\,F(k,\kh)=1,
 \label{eq:Fdef}
\end{equation}
from its zero-monopole anisotropic part $Q$.  The measure $F$ may be a smooth function or a distribution; for a compact torus it is the windowed directional spectral measure on the allowed lattice modes.  Its nonmonopole moments are
\begin{equation}
 \begin{aligned}
 q_{LM}(k)&\equiv\int\dd\Omega_{\kh}\,Q(k,\kh)Y^*_{LM}(\kh)\\
 &=\int\dd\Omega_{\kh}\,F(k,\kh)Y^*_{LM}(\kh),\qquad L>0.
 \end{aligned}
 \label{eq:qdef}
\end{equation}
with $q^*_{LM}=(-1)^Mq_{L,-M}$.

The same $q_{LM}$ are the fractional anisotropy multipoles of the present-day tensor energy density, provided the tensor transfer
is direction independent:
\begin{equation}
 \Om(k,\kh)=\bar\Omega_{\rm GW}(k)F(k,\kh),
 \label{eq:omegaq}
\end{equation}
where, in the usual short-wavelength and oscillation-averaged convention,
\begin{equation}
 \bar\Omega_{\rm GW}(k)=\frac{1}{12}\left(\frac{k}{a_0H_0}\right)^2T_h^2(k)P_h(k).
 \label{eq:omegabar}
\end{equation}
We define the absolute anisotropy multipoles by
\begin{equation}
 \begin{aligned}
 \Omega_{LM}(k)&\equiv\int\dd\Omega_{\kh}\,Y^*_{LM}(\kh)\\
 &\quad\times\left[\Om(k,\kh)-\bar\Omega_{\rm GW}(k)\right],\qquad L>0.
 \end{aligned}
 \label{eq:OmegaLMdef}
\end{equation}
Hence
\begin{align}
 \Omega_{LM}(k)&=\bar\Omega_{\rm GW}(k)\,q_{LM}(k),\nonumber\\
 P_h(k)\,q_{LM}(k)&=\frac{12a_0^2H_0^2}{k^2T_h^2(k)}\Omega_{LM}(k).
 \label{eq:omega-to-q}
\end{align}
The primary source entering the CMB calculation is $P_h(k)q_{LM}(k)$; the relation to absolute energy density is quoted for the standard regime in which Eq.~\eqref{eq:omegabar} applies.

\section{\texorpdfstring{$B$}{B}-mode transfer kernel}

\begin{figure*}[t]
 \centering
 \includegraphics[width=0.98\textwidth]{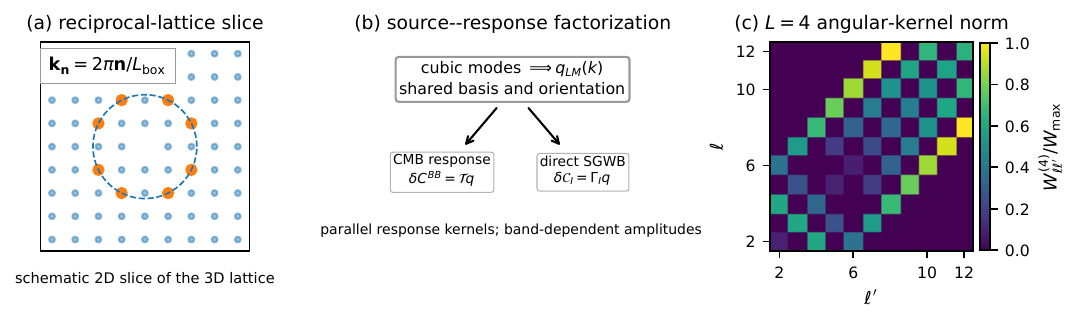}
 \caption{Source--response interpretation of the tensor topology signal.  Panel (a) is a schematic two-dimensional slice of the three-dimensional reciprocal lattice.  Panel (b) separates the topology-generated $q_{LM}$ source from the parallel CMB and direct-SGWB response kernels; neither response is an input to the other.  Panel (c) shows the calculated normalized angular-kernel block norm $\widehat W^{(4)}_{\ell\ell'}\equiv W^{(4)}_{\ell\ell'}/\max W^{(4)}_{\ell\ell'}$, where $W^{(4)}_{\ell\ell'}=[\sum_{Mmm'}|\Kern^{BB;4M}_{\ell m,\ell'm'}|^2]^{1/2}$.  The colors are kernel norms, not forecast weights.}
 \label{fig:kernel}
\end{figure*}

Using the standard all-sky polarization convention and spin-weighted spherical harmonics \cite{Goldberg1967}, the tensor contribution to the CMB $B$-mode coefficient is
\begin{equation}
 a^B_{\ell m}=4\pi(-\ii)^\ell\sum_{\lambda=\pm1}\int\frac{\dd^3 k}{(2\pi)^3}
 h_\lambda(\mathbf k)\Delta^B_\ell(k)\epsilon_B^\lambda
 {}_{-2\lambda}Y^*_{\ell m}(\kh),
 \label{eq:almB}
\end{equation}
where
\begin{equation}
 \epsilon_B^\lambda=\ii\lambda .
\end{equation}
Different spin-harmonic or partial-wave phase conventions only rephase covariance blocks; the invariant norms and selection rules below are unchanged.  Substituting Eq.~\eqref{eq:tensor-stat} into $\langle a^B_{\ell m}a^{B*}_{\ell'm'}\rangle$ gives the isotropic diagonal spectrum plus the anisotropic part.  Since
\begin{equation}
 \int\frac{\dd^3 k}{(2\pi)^3}\frac{2\pi^2}{k^3}=\frac{1}{4\pi}\int\dd\ln k\,\dd\Omega_{\kh},
\end{equation}
the anisotropic covariance is
\begin{widetext}
\begin{align}
 \delta C^{BB}_{\ell m,\ell'm'}
 &=4\pi\ii^{\ell'-\ell}\int\dd\ln k\,\frac{P_h(k)}{2}
 \Delta^B_\ell(k)\Delta^B_{\ell'}(k)
\nonumber\\
 &\quad\times\sum_{\lambda=\pm1}\sum_{LM}q_{LM}(k)
 \int\dd\Omega_{\kh}\,Y_{LM}(\kh)
 {}_{-2\lambda}Y^*_{\ell m}(\kh)
 {}_{-2\lambda}Y_{\ell'm'}(\kh).
 \label{eq:deltaC-raw}
\end{align}
Define the helicity-averaged spin-weighted Gaunt kernel
\begin{equation}
 \Kern^{BB;LM}_{\ell m,\ell'm'}\equiv\frac12\sum_{\lambda=\pm1}
 \int\dd\Omega_{\kh}\,Y_{LM}(\kh)
 {}_{-2\lambda}Y^*_{\ell m}(\kh)
 {}_{-2\lambda}Y_{\ell'm'}(\kh).
 \label{eq:Kdef}
\end{equation}
Then
\begin{equation}
 % \boxed{\;
 \delta C^{BB}_{\ell m,\ell'm'}=
 \sum_{LM}\int\dd\ln k\,
 \Trans^{BB;LM}_{\ell m,\ell'm'}(k)q_{LM}(k)
 % \;}
 \label{eq:kernelmap}
\end{equation}
with
\begin{equation}
 % \boxed{\;
 \Trans^{BB;LM}_{\ell m,\ell'm'}(k)=4\pi\ii^{\ell'-\ell}
 P_h(k)\Delta^B_\ell(k)\Delta^B_{\ell'}(k)
 \Kern^{BB;LM}_{\ell m,\ell'm'} .
 % \;}
 \label{eq:Tkernel}
\end{equation}
Equations~\eqref{eq:kernelmap} and \eqref{eq:Tkernel} give the transfer kernel from primordial SGWB anisotropy to CMB $B$-mode covariance.  In terms of $\Omega_{LM}$, Eq.~\eqref{eq:omega-to-q} recasts this as
\begin{align}
 \delta C^{BB}_{\ell m,\ell'm'}
 &=\sum_{LM}\int\dd\ln k\,
 4\pi\ii^{\ell'-\ell}\frac{12a_0^2H_0^2}{k^2T_h^2(k)}\nonumber\\
 &\quad\times \Delta^B_\ell(k)\Delta^B_{\ell'}(k)
 \Kern^{BB;LM}_{\ell m,\ell'm'}\Omega_{LM}(k).
 \label{eq:Omega-kernel}
\end{align}
This form should be interpreted as the ultra-low-$k$ primordial SGWB projected into the CMB, not as a transfer from a nanohertz or millihertz stochastic background.

The angular kernel has a closed form.  With ${}_sY^*_{\ell m}=(-1)^{m+s}{}_{-s}Y_{\ell,-m}$ and the spin-weighted Gaunt integral,
\begin{align}
 \Kern^{BB;LM}_{\ell m,\ell'm'}&=(-1)^m
 \left[\frac{(2L+1)(2\ell+1)(2\ell'+1)}{4\pi}\right]^{1/2}\nonumber\\
 &\quad\times
 \begin{pmatrix}L&\ell&\ell'\\ M&-m&m'\end{pmatrix}
 \begin{pmatrix}L&\ell&\ell'\\ 0&-2&2\end{pmatrix}
 \frac{1+(-1)^{L+\ell+\ell'}}{2} .
 \label{eq:K3j}
\end{align}
\end{widetext}
Thus $\delta C^{BB}_{\ell m,\ell'm'}$ can be nonzero only if
\begin{equation}
 |\ell-\ell'|\le L\le \ell+\ell',\qquad M=m-m',
 \qquad L+\ell+\ell'\;{\rm even}.
 \label{eq:BBselect}
\end{equation}
For $XB$ with $X=T,E$, the helicity factor is instead proportional to $\lambda$, and the parity projector in Eq.~\eqref{eq:K3j} is replaced by $[1-(-1)^{L+\ell+\ell'}]/2$.  Therefore $TB$ and $EB$ require $L+\ell+\ell'$ odd.  For parity-even compact topologies such as the cubic $T^3$, the SGWB source multipoles have even $L$, so diagonal $TB$ and $EB$ remain forbidden while off-diagonal blocks with $\ell+\ell'$ odd can be topology-allowed.
Although pointwise reconstruction of a delta-supported angular measure requires arbitrarily high $L$, a band-limited CMB covariance does not.  The triangle condition gives the exact closure
\begin{equation}
 L_q^{\max}=2\ell_{\max}.
 \label{eq:exactclosure}
\end{equation}
Thus the $\ell_{\max}=12$ calculation closes at $L_q^{\max}=24$.  Independent direct-shell and $q_{LM}$--Gaunt calculations agree at this closure to relative Frobenius residuals $1.4\times10^{-14}$--$3.0\times10^{-14}$ for the tested cubic shells and orientations; at $L_q^{\max}=20$ the residuals remain $0.18$--$0.25$.  Figure~\ref{fig:kernel} summarizes the source--response construction.

\section{Compact-topology source multipoles}

For a cubic three-torus,

\begin{equation}
 \mathbf k_{\mathbf n}=\frac{2\pi}{L_{\rm box}}\mathbf n,
 \qquad \mathbf n\in\mathbb Z^3\setminus\{0\} .
 \label{eq:t3modes}
\end{equation}
In a radial window, the normalized directional tensor-power spectral measure can be written
\begin{equation}
 F(k,\kh)=\frac{4\pi}{N_W(k)}\sum_{\mathbf n\ne0}w_{\mathbf n}(k)
 \delta^{(2)}\!\left(\kh-\hat{\mathbf k}_{\mathbf n}\right),
 \label{eq:Ftorus}
\end{equation}
where $N_W(k)=\sum_{\mathbf n}w_{\mathbf n}(k)>0$ and the non-negative weights $w_{\mathbf n}$ encode the radial window, source power, and any smooth shell weighting.  Since the monopole is $\sqrt{4\pi}Y_{00}=1$, the nonmonopole coefficients are
\begin{equation}
 % \boxed{
 q^{T^3}_{LM}(k)=\frac{4\pi}{N_W(k)}\sum_{\mathbf n\ne0}w_{\mathbf n}(k)
 Y^*_{LM}\!\left(\hat{\mathbf k}_{\mathbf n}\right),\qquad L>0.
 % }
 \label{eq:qT3}
\end{equation}
The angular normalization divides out shell multiplicity, which remains in the radial monopole measure.  For an exact shell $s$ with multiplicity $N_s$, volume $V=L_{\rm box}^3$, and wave number $k_s$, the factorized integral uses
\begin{equation}
 [P_h(k)\dd\ln k]_s=\frac{2\pi^2N_s}{Vk_s^3}P_h^{\rm cov}(k_s)
 \delta(\ln k-\ln k_s)\dd\ln k,
 \label{eq:shellmeasure}
\end{equation}
where $P_h^{\rm cov}$ is the covering-space spectrum entering the compact Fourier-mode covariance.  With Eq.~\eqref{eq:shellmeasure}, substituting Eq.~\eqref{eq:qT3} into Eq.~\eqref{eq:kernelmap} is exactly equivalent to the direct compact-mode sum.

For complete cubic shells, $F(k,\kh)$ is invariant under the octahedral group $O_h$.  Hence $q^{T^3}_{LM}$ vanishes unless the $SO(3)$ representation $D^{(L)}$ contains the fully symmetric cubic irrep $A_{1g}$ \cite{Hamermesh1962,Riazuelo2004}.  The first nonmonopole allowed multipoles are
\begin{equation}
 L=4,6,8,10,12,\ldots,
\end{equation}
with nonmonotonic multiplicities; in particular $m_{12}(A_{1g})=2$, $m_{14}(A_{1g})=1$, and $m_{16}(A_{1g})=2$.  The corresponding amplitudes are not fixed by symmetry: they vary with the radial shell and with the smooth weights chosen for the window.  Symmetry alone pins down which multipoles can appear; whether they are detectable depends separately on their amplitude and on the instrument response.

The single-wavevector $q_{LM}(k)$ representation extends directly to statistically homogeneous, wavevector-diagonal cases, including general three-tori and partially compact pure-translation slab or chimney spaces, with the corresponding reciprocal lattices and point-group restrictions.  In inhomogeneous quotients, topology can correlate distinct covering-space wavevectors.  The source--response factorization still generalizes, but the source object is then a covariance matrix carrying wavevector and polarization indices, or equivalently a double-wavevector covariance $\Xi_{hh'}(\mathbf k,\mathbf k')$, rather than a single $q_{LM}(k)$.

\subsection{Cross-frequency interpretation}

Equation~\eqref{eq:qT3} carries a cross-band reading as well.  The coefficients $q_{LM}^{T^3}(k)$ are not in general frequency independent, since each frequency bin picks out a different set of lattice shells and weights.  What the geometry does fix is the angular subspace itself and its orientation.  Concretely,

\begin{equation}
 q_{LM}^{T^3}(k;R)=\sum_{r=1}^{m_L(A_{1g})}A_{Lr}(k)
 c_{LM}^{(r),{\rm cubic}}(R),
 \label{eq:crossfreq}
\end{equation}
where $R\in SO(3)$ is the orientation of the fundamental cube, $c_{LM}^{(r),{\rm cubic}}$ are frequency-independent cubic harmonics, and the amplitudes $A_{Lr}(k)$ absorb the shell, window, and source physics.  What topology predicts, then, is a shared angular basis and orientation across bands rather than a single frequency-independent amplitude.  If the amplitudes are left completely free, the shared subspace tests cubic symmetry and orientation but does not determine $L_{\rm box}$ without radial-spectrum information or resolved mode spacing.

The CMB kernel of Eq.~\eqref{eq:kernelmap} probes the ultra-low-$k$ end of this template.  A direct SGWB experiment samples the same source multipoles through its own time-dependent response.  For a detector baseline or TDI-channel pair $I$, the expectation value of an anisotropy-sensitive cross-spectral observable can be written schematically as
\begin{equation}
 \delta \mathcal C_I(f,t)\propto \bar\Omega_{\rm GW}(f)
 \sum_{LM}\Gamma_I^{LM}(f,t)q_{LM}(f),
 \label{eq:directresponse}
\end{equation}
where $\Gamma_I^{LM}$ is the generalized overlap-reduction or antenna-response kernel.  LISA and Taiji/LISA studies already cast their sensitivity in terms of SGWB anisotropy multipoles \cite{Bartolo2022LISAAnisotropy,ZhaoWang2024TaijiLISA}.  Equation~\eqref{eq:crossfreq} imposes the additional topology constraint that the allowed angular subspace and its orientation coincide with those of the CMB tensor channel.  Astrophysical foregrounds and local source anisotropies need not respect this template and must be modeled separately.  For horizon-scale compactification the CMB $B$-mode channel is expected to be a particularly clean low-frequency tensor probe, since it samples $k\sim \ell/\chi_*$ where lattice averaging is weakest.  At high $k$, dense shell sampling can suppress topology-induced amplitudes through directional averaging; the LISA, Taiji, TianQin, and PTA bands therefore provide complementary consistency tests, rather than guaranteed high-sensitivity channels, if a cosmological SGWB component exists \cite{LISA2017,Taiji2021,TianQin2016}.

For comparison, transverse-traceless tensor harmonics on $S^3$ carry the Laplacian label
\begin{equation}
 q_n^2=\frac{n^2-3}{R^2},\qquad n\ge3,
\end{equation}
whereas the dynamical tensor frequency in the closed-FLRW wave equation contains the curvature shift
\begin{equation}
 K_n^2=\frac{n^2-1}{R^2} .
\end{equation}
The degeneracy is $d_n=2(n^2-4)$ \cite{RubinOrdonez1984,Tomita1982}.  Since $S^3$ is maximally symmetric, it supplies no deterministic preferred-direction template in the ensemble average.  A finite-mode realization has rms multipole amplitude scaling as $N_{\rm modes}^{-1/2}$ and angular-power ratio scaling as $\langle C_L/C_0\rangle_{\rm fm}\sim N_{\rm modes}^{-1}\propto R^{-3}$ in narrow physical bins.
\begin{figure*}[t]
 \centering
 \includegraphics[width=0.98\textwidth]{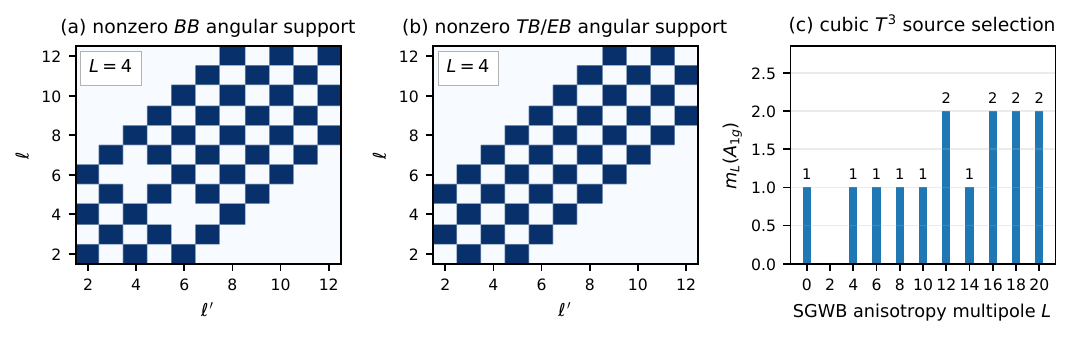}
 \caption{Selection rules for the SGWB-anisotropy transfer kernel.  For $L=4$, panels (a) and (b) show nonzero block support after imposing the triangle condition, the spin-row Wigner coefficient, and the appropriate parity projector, with $\ell,\ell'\ge2$.  The $BB$ helicity average contains $[1+(-1)^{L+\ell+\ell'}]/2$, whereas $TB/EB$ contains $[1-(-1)^{L+\ell+\ell'}]/2$.  A colored block indicates that at least one $(M,m,m')$ element is nonzero; it does not imply equal amplitude.  Panel (c) gives exact $A_{1g}$ multiplicities for a parity-even cubic $T^3$ through $L=20$.}
 \label{fig:selection}
\end{figure*}

\section{Forecast diagnostics}

For a topology template define the rotationally invariant anisotropic part

\begin{align}
 \delta C&=C-\Pi_{\rm iso}(C),\nonumber\\
 \Pi_{\rm iso}(C)^{XY}_{\ell m,\ell'm'}&=
 \delta_{\ell\ell'}\delta_{mm'}\frac{1}{2\ell+1}
 \sum_{\mu=-\ell}^{\ell}C^{XY}_{\ell\mu,\ell\mu}.
 \label{eq:projection}
\end{align}
For a single-amplitude template, the ideal full-sky Gaussian-covariance diagnostic is \cite{Tegmark1997}
\begin{equation}
 (S/N)^2=\frac12\Tr\left(C_0^{-1}\delta C\,C_0^{-1}\delta C\right),
 \label{eq:fisher}
\end{equation}
where $C_0$ is the statistically isotropic CAMB covariance for the selected fields, evaluated without the compact-space radial cutoff.  We use CAMB 1.6.5 transfer functions \cite{Lewis2000}, Planck-like parameters \cite{PlanckParams2020}, and $r=0.01$ \cite{BICEPKeck2021}.  The compact covariance is evaluated by a direct cubic-lattice sum with $V=L_{\rm box}^3$ and $x_{\max}\equiv k_{\max}\chi_*=50$.  The calculation is full sky, uses a fixed aligned cube and fixed cosmological parameters, adopts a Gaussian covariance approximation, assumes no instrumental noise or foreground covariance, and includes no marginalization or scale/orientation look-elsewhere penalty.  We therefore call it an idealized fixed-template diagnostic rather than a mission-level limit.

\begin{figure*}[t]
 \centering
 \includegraphics[width=0.98\textwidth]{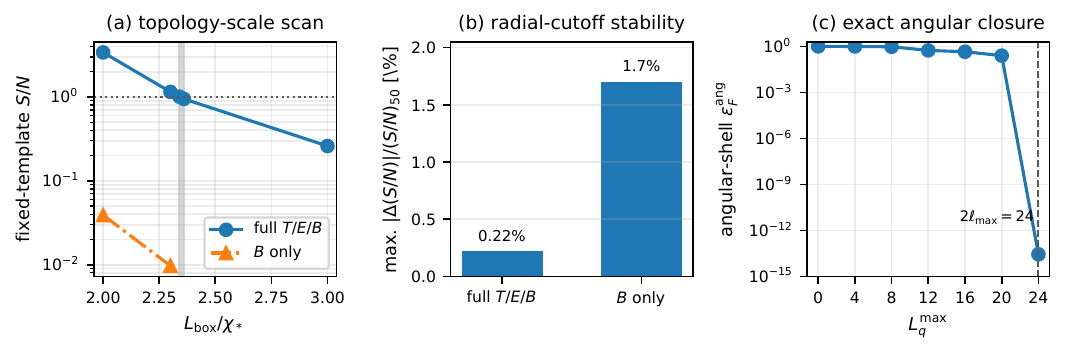}
 \caption{Full-sky fixed-template diagnostics for $2\le\ell\le12$ in the Gaussian covariance approximation, with a fixed aligned cube, fixed cosmological parameters, $r=0.01$, no instrumental-noise or foreground covariance, and no parameter marginalization or scale/orientation look-elsewhere correction.  (a) Topology-scale scan evaluated from the direct compact-mode sum at $x_{\max}=50$; the gray strip brackets the sampled-grid combined $S/N=1$ transition between $L_{\rm box}/\chi_*=2.34$ and 2.36.  (b) Maximum fractional radial-cutoff variation over $40\le x_{\max}\le80$ at $L_{\rm box}/\chi_*=2.3$, where the combined statistic exceeds unity.  (c) Single-shell closure test comparing the direct compact-mode sum with the factorized $q_{LM}$--Gaunt construction; exact band closure occurs at $L_q^{\max}=2\ell_{\max}=24$.}
 \label{fig:diagnostics}
\end{figure*}

Figure~\ref{fig:diagnostics}(a) shows the topology-scale scan for $2\le\ell\le12$.  The combined $T/E/B$ values are $S/N=3.394$ at $L_{\rm box}/\chi_*=2$, $1.149$ at 2.30, $1.007$ at 2.34, $0.948$ at 2.36, and $0.260$ at 3.00.  On the sampled grid,
\begin{equation}
 2.34<\left(L_{\rm box}/\chi_*\right)_{S/N=1}<2.36.
 \label{eq:crossing}
\end{equation}
The tensor-only statistic remains far below unity across this interval and therefore supplies an independent consistency channel rather than a competing detection threshold.  The $B$-only values are $S/N=0.039$ at $L_{\rm box}/\chi_*=2$ and $0.0097$ at 2.3.

At $L_{\rm box}/\chi_*=2.3$, changing $x_{\max}$ from 40 to 80 changes the combined statistic by at most $0.22\%$ relative to $x_{\max}=50$ and the $B$-only value by at most $1.7\%$.  Figure~\ref{fig:diagnostics}(c) independently compares a direct angular shell sum with the $q_{LM}$--Gaunt contraction for the complete cubic shell $|\mathbf n|^2=5$: the relative matrix residual reaches $2.8\times10^{-14}$ at $L_q^{\max}=24$, and the Hermiticity residual across the tested shells and orientations does not exceed $1.6\times10^{-16}$.

Direct summation is suitable for a single sparse low-$\ell$ shell.  The factorized form instead permits the same topology-dependent source multipoles to be reused across transfer functions, detector responses, orientations, source spectra, and frequency bands.

\section{Discussion}

Equations~\eqref{eq:kernelmap}--\eqref{eq:K3j} are the central result: CMB $B$-mode covariance acts as a linear transfer map from primordial SGWB anisotropy onto an observable spin-2 covariance.  This does not supplant the compact-topology covariance formalisms, which deliver the full correlation-matrix endpoint; rather, it isolates the intermediate SGWB anisotropy multipoles that compact tensor eigenmodes generate and that the endpoint formalisms leave implicit.  The direct-shell closure test confirms that the factorized and direct representations agree to numerical precision when $L_q^{\max}=2\ell_{\max}$.

Equation~\eqref{eq:crossfreq} defines a cross-frequency consistency relation.  Compact topology does not predict identical numerical $q_{LM}$ amplitudes at CMB, PTA, and space-interferometer frequencies; what it predicts is a shared angular subspace, a shared orientation, and shared selection rules, with the amplitudes left band-dependent.  The CMB $B$-mode covariance is then the low-frequency, transfer-filtered tensor channel, and direct SGWB experiments probe the same underlying geometry through their own response kernels.  Accordingly, CMB polarization and PTA, LISA, Taiji, and TianQin anisotropy searches test a common geometric template through distinct response kernels.

For a cubic $T^3$ the first nontrivial SGWB anisotropy components are $L=4,6,8,\ldots$, and the $B$-mode covariance inherits the Wigner-symbol selection rules of the transfer kernel.  The same machinery supplies a matched-template statistic: writing a topology template as $q_{LM}(k;\Theta_{\rm top})=A\,u_{LM}(k;\Theta_{\rm top})$, Eq.~\eqref{eq:kernelmap} fixes $S^{BB}$ and Eq.~\eqref{eq:fisher} returns the Fisher information for the amplitude $A$.

These fixed-template diagnostics are deliberately idealized and should be regarded as optimistic sensitivity estimates.  Even under these assumptions, the tensor $B$-mode route is not on its own a strong near-term detection channel in the cubic models tested once $L_{\rm box}$ exceeds the last-scattering diameter.  Scalar $T/E$ covariance is not itself an SGWB signal, yet it belongs to the same compact-eigenmode boundary-value problem and holds most of the practical CMB topology information.  A realistic search will want to combine scalar and tensor covariance blocks while retaining the $B$ modes as a conceptually clean primordial-tensor consistency channel.  A mission-grade treatment would include realistic masks and foreground covariance, pure-$B$ construction, a specified lensing/delensing treatment including connected non-Gaussian covariance, and marginalization over the compactification scale and orientation.  Beyond that, a full closed-universe $S^3$ transfer calculation, the lower-symmetry compact quotients, and joint CMB/direct-SGWB template fits are natural extensions.

\begin{acknowledgments}
This work is supported by the National Key R\&D Program of China under Grants No. 2025YFE0217300 and No. 2024YFC2206902.
\end{acknowledgments}

\bibliographystyle{apsrev4-2}
\bibliography{cmb_topology_refs_revision1_4}
\end{document}